\documentclass[conference]{IEEEtran}
\IEEEoverridecommandlockouts

\usepackage{flushend}
\usepackage{cite}
\usepackage{amsmath,amssymb,amsfonts}
\usepackage{algorithmic}
\usepackage{algorithm}
\usepackage{graphicx}
\usepackage{textcomp}
\usepackage{xcolor}
\usepackage{booktabs}
\usepackage{subcaption}
\usepackage{overpic}
\usepackage{adjustbox}
\usepackage{tabularx}
\usepackage{siunitx}
\usepackage{float}
\DeclareMathOperator*{\argmin}{argmin} 
\def\BibTeX{{\rm B\kern-.05em{\sc i\kern-.025em b}\kern-.08em
    T\kern-.1667em\lower.7ex\hbox{E}\kern-.125emX}}
\begin{document}

\title{MR2US-Pro: Prostate MR to Ultrasound Image Translation and Registration Based on Diffusion Models
}

\author{
Xudong Ma\textsuperscript{1},
Nantheera Anantrasirichai\textsuperscript{1},
Stefanos Bolomytis\textsuperscript{2}, 
Alin Achim\textsuperscript{1}\\
\textsuperscript{1}Visual Information Laboratory, University of Bristol, Bristol, UK \\
\textsuperscript{2}Southmead Hospital, North Bristol NHS Trust, UK \\
\{xudong.ma, n.anantrasirichai, alin.achim\}@bristol.ac.uk, stefanos.bolomytis@nbt.nhs.uk
}

\maketitle

\begin{abstract}
The diagnosis of prostate cancer increasingly depends on multimodal imaging, particularly magnetic resonance imaging (MRI) and transrectal ultrasound (TRUS). However, accurate registration between these modalities remains a fundamental challenge due to the differences in dimensionality and anatomical representations. In this work, we present a novel framework that addresses these challenges through a two-stage process: TRUS 3D reconstruction followed by cross-modal registration. Unlike existing TRUS 3D reconstruction methods that rely heavily on external probe tracking information, we propose a totally probe-location-independent approach that leverages the natural correlation between sagittal and transverse TRUS views. With the help of our clustering-based feature matching method, we enable the spatial localization of 2D frames without any additional probe tracking information. For the registration stage, we introduce an unsupervised diffusion-based framework guided by modality translation. Unlike existing methods that translate one modality into another, we map both MR and US into a pseudo intermediate modality. This design enables us to customize it to retain only registration-critical features, greatly easing registration. To further enhance anatomical alignment, we incorporate an anatomy-aware registration strategy that prioritizes internal structural coherence while adaptively reducing the influence of boundary inconsistencies. Extensive validation demonstrates that our approach outperforms state-of-the-art methods by achieving superior registration accuracy with physically realistic deformations in a completely unsupervised fashion.
\end{abstract}


\section{Introduction}

Prostate cancer remains a significant global health concern, representing the second most frequently diagnosed malignancy in men worldwide \cite{bratt2024population}. The clinical management of this disease increasingly depends on advanced imaging technologies to achieve precise diagnosis and guide therapeutic interventions. Among these modalities, magnetic resonance imaging (MRI) has emerged as the gold standard for detailed anatomical visualization due to its unparalleled soft tissue contrast resolution, particularly in identifying suspicious lesions \cite{weinreb2016pi}. In addition, transrectal ultrasound (TRUS) is also widely used due to its real-time imaging capabilities, cost-effectiveness, and established role in procedural guidance during biopsy \cite{valerio2014role}.  

The integration of magnetic resonance (MR) and ultrasound (US) image through accurate registration has significant clinical value, but presents fundamental technical challenges \cite{ma2024pmt}. One key obstacle is the dimensional disparity between 3D MR volumes and 2D US videos, which requires volumetric reconstruction of US data before registration. Conventional 3D ultrasound reconstruction methods often rely on external probe tracking information to establish spatial relationships between frames \cite{daoud2015freehand,guo2022ultrasound}.This introduces additional hardware requirements and procedural complexity.

Therefore, in this paper, we propose a novel approach that does not involve any probe tracking for 3D reconstruction of prostate US images. Our method stitches sagittal US frames into a 2D map, which serves as a reference for deriving the relative positions of every frame. We perform this stitching first, rather than relying directly on consecutive frame comparisons, as the latter can easily propagate errors across subsequent frames. By exploiting the orthogonal relationship between sagittal and transverse images, we transfer the spatial relationships between sagittal frames to transverse frames and construct a 3D volume.

Beyond dimensional disparities, the  modality-specific discrepancies between MR and US images still challenge the traditional registration methods. Most of these methods typically assume consistent intensity or texture relationships. Their usage of similarity metrics (e.g., mutual information) often fail due to nonlinear intensity relationships between MR and US. Alternatively, some methods involve anatomical structure segmentation before registration to alleviate the modality gap. However, these methods depend on large amount of expert annotated data, limiting their scalability and clinical applicability.

As a result, we introduce a novel unsupervised diffusion-based framework that fundamentally rethinks this problem through registration-oriented modality translation. Unlike conventional translation methods that focus on bidirectional image conversion (\textnormal{MRI$\leftrightarrow$US}) with excessive emphasis on visual fidelity, we propose a hierarchical feature disentanglement approach that customizes the transformation of MR and US images into an anatomically coherent intermediate modality. This is achieved through two complementary mechanisms. On one hand, we use shallow-layer features of a diffusion network with larger convolutional kernels (7×7) to ensure the translated images maintain high consistency in texture and prostate internal anatomical features. On the other hand, we employ deep-layer features with smaller kernels (3×3) to make sure the translated results preserve the essential boundaries of the corresponding input images properly. 




Although our modality translation method achieves anatomical coherence of the prostate’s internal regions, the inherent differences in imaging principles still lead to varying boundary thickness and morphology. Existing registration methods struggle with this challenge, because they either treat all voxels equally or focus on high-information regions (boundary areas). This usually results in over-registration of the boundaries, which does not meet anatomical needs. To address this, we propose a novel anatomy-aware diffusion model for registration. In our approach, voxel importance decreases with its information content. Higher weights are assigned to the low-information yet coherent internal regions of the prostate, while the influence of high-information boundary areas is downweighted. This mechanism ensures that the deformation field prioritizes aligning the prostate interior and treats boundary variations as lower-confidence guidance. As a result, our method achieves more anatomically and clinically accurate registration.


Building upon this design paradigm, our principal contributions can be summarized as follows:
\begin{enumerate}
    \item We establish the first comprehensive pipeline, called MR2US-Pro, for cross-dimensional and cross-modal registration from prostate MR to US images that directly starts from raw clinical data - 2D US video streams and 3D MR volumes;
    \item We propose an innovative approach for 3D reconstruction of prostate TRUS images that doesn't rely on  any external probe location information throughout the whole process.
    \item We introduce a modality translation solution to fundamentally reformulate multimodal registration into a more tractable monomodal alignment problem;
    \item We pioneer the concept of a customized pseudo-intermediate modality, thereby addressing the bottlenecks of existing modality-conversion methods. Leveraging an anatomically coherent modality translation (ACMT) scheme, we strategically retain boundary structures while homogenizing modality-specific characteristics within the prostate interior, effectively narrowing cross-modal gaps without sacrificing essential anatomy. Consequently, the approach provides more favorable conditions for accurate and robust registration.
    \item Our anatomy-aware registration network subsequently concentrates the alignment energy on the coherent prostate internal structures of the intermediate modality, thereby enhancing the registration outcome.
\end{enumerate}


The remainder of this paper is organized as follows: Section 2 reviews existing methods in 3D US reconstruction, modality translation, and registration. Section 3 presents the details of our proposed framework. Section 4 reports both qualitative and quantitative results for both modality translation and registration. Finally, Section 5 summarizes the ideas and contributions of our work.


\section{Related Work}
\subsection{Prostate TRUS 3D Reconstruction}
Traditional prostate TRUS 3D reconstruction methods typically rely on external tracking systems, such as electromagnetic or optical trackers, to capture the probe’s spatial position and orientation during scanning. This information is then used to align 2D frames into a 3D volume \cite{wen2013accurate,daoud2015freehand,hafizah2010development,rohling1999comparison}. While effective, these systems increase complexity and cost, limiting their clinical adoption.

More recently, deep learning has enabled methods that estimate probe motion directly from consecutive frames\cite{wein2020three,guo2022ultrasound}. However, they still require ground truth probe motion data for training, which is difficult to obtain in clinical environments due to the lack of tracking equipment and publicly available datasets. These limitations highlight the need for a probe-location-independent solution. 


\begin{figure*}
    \centering
    \includegraphics[width=0.75\textwidth]{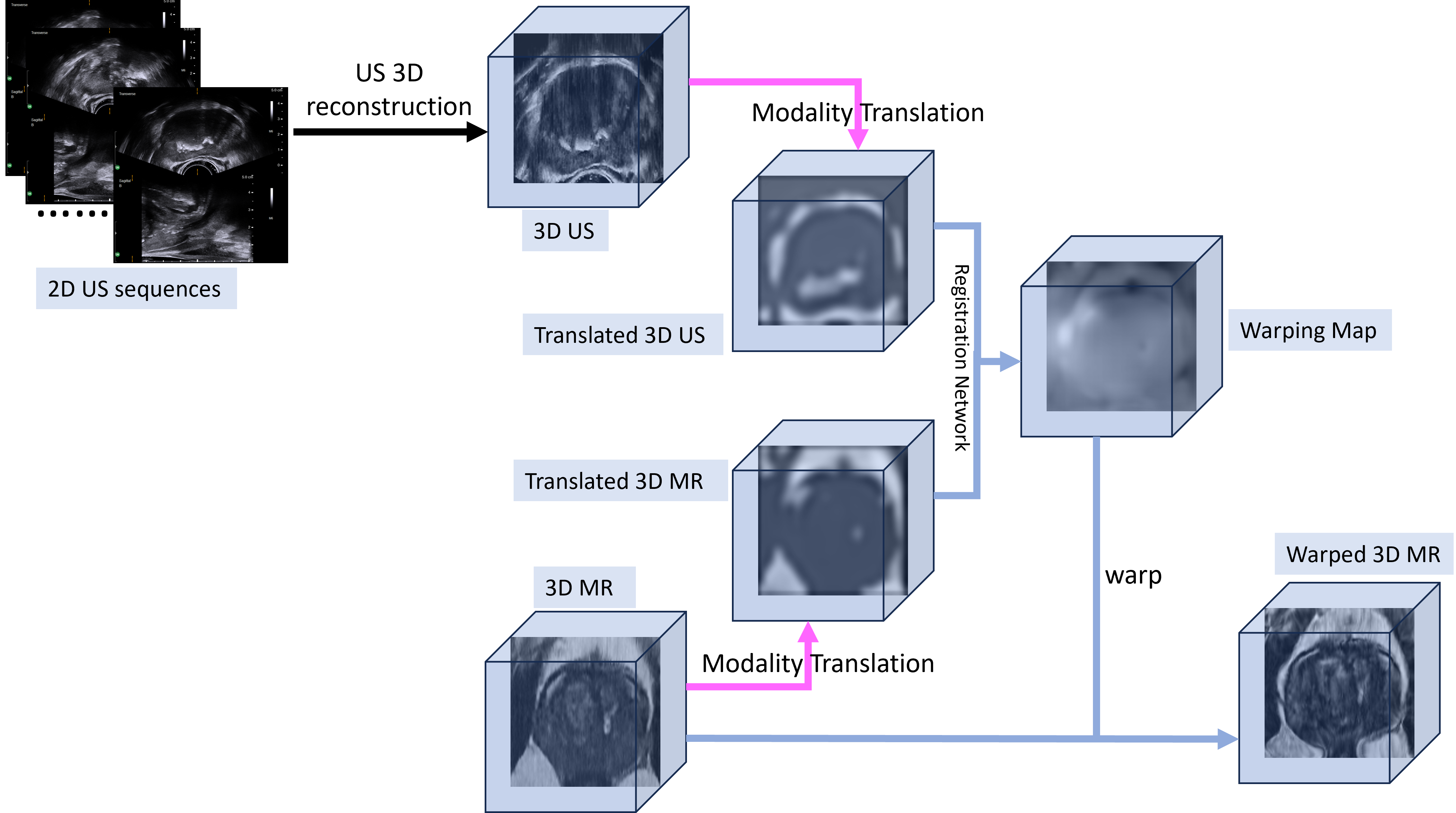}
    \caption{The end-to-end workflow of our MR2US-Pro}
    \label{fig:end-to-end}
\end{figure*}

\subsection{Modality Translation} \label{sec:MT_review}
Even with proper 3D US reconstruction, registering prostate 3D MR and US images remains a longstanding challenge due to their significant differences in imaging physics, contrast, and anatomical appearance \cite{weinreb2016pi, valerio2014role}. Modality translation has emerged as a promising solution for bridging the representation gap between MR and US, enabling registration in a common feature space. 

Some works leveraged generative adversarial networks (GANs) \cite{tian2024oct2confocal,liu2017unsupervised,jiao2020self} for direct modality translation. While these methods have demonstrated visually realistic synthesis, they often fail to preserve fine anatomical details critical for registration. Meanwhile, these methods unavoidably suffer from inherent limitations of GAN-based frameworks, such as unstable training dynamics, slow convergence, and susceptibility to mode collapse. Although the recent diffusion model-based modality translation methods\cite{ho2020denoising,wolleb2022diffusion,kim2023unpaired} iteratively refine noisy inputs, leading to superior feature consistency, they still mainly focus on converting images from one modality to another and improving visual fidelity rather than addressing anatomical consistency. This limitation becomes particularly critical in registration tasks, where anatomical representation differences caused by modality variations can significantly compromise performance.

Our earlier work, PMT\cite{ma2024pmt}, attempted to mitigate this issue by incorporating shallow layer cross-modal texture similarity constraints into the diffusion process, guiding the translation toward an intermediate modality with improved texture alignment. While PMT achieved better registration results, it still preserved too much image details due to the emphasis on image realism, particularly in the internal prostate areas. Therefore, it does not truly a customized modality design.


To address these limitations, we propose our ACMT method to integrate the anatomical features of both MR and US images in a truly customized fashion, achieving minimal modality differences while only retaining the essential boundary features.

\subsection{Image Registration}
Intensity-based methods are among the most commonly used approaches for image registration. They typically rely on image similarity metrics such as mutual information (MI) and normalized cross-correlation (NCC) \cite{tascon2022accuracy} to guide the alignment process \cite{pluim2003mutual,rueckert1999nonrigid}. However, such methods are not suitable for our cross-modality registration task, as the intensity distributions between different modalities exhibit highly non-linear relationships. This makes it difficult for traditional similarity measures to accurately capture the underlying anatomical correspondences. Although our ACMT method creates a modality where the internal prostate areas typically appear as coherent low-intensity areas, high-intensity boundaries still exhibit inconsistent thickness and morphology across images.

Feature-based registration methods, which leverage anatomical landmarks or segmentation masks, have been proposed as an alternative \cite{chen2021mr, jiang2023segmentation}. Although they can effectively circumvent the issue of modality discrepancies, they typically require large annotated datasets for training. This highly limits their applicability in real-world clinical settings where labeled data is scarce.

Recently, unsupervised methods have been developed, such as VoxelMorph \cite{balakrishnan2019voxelmorph} and DiffusionMorph \cite{kim2022diffusemorph}, which primarily rely on intensity-based similarity metrics to learn deformation fields without ground-truth labels. These methods, however, treat all voxels equally, applying uniform weighting across the image domain without distinguishing anatomical regions. While newer approaches like FSDiffReg \cite{qin2023fsdiffreg} incorporate voxel-wise weighting schemes, these methods still tend to place more emphasis on regions with higher intensities, which in our case, often results in an overemphasis on boundary areas that exhibit significant variability. Therefore, a registration strategy that more explicitly accounts for anatomical semantics is required to address our problem effectively.

\section{Proposed Framework}
In this section, we provide a comprehensive explanation of the workflow for our proposed MR2US-Pro. As shown in Fig.~\ref{fig:end-to-end}, we first perform 3D reconstruction on 2D US sequences using our probe-location-independent method. Then, both the 3D MR and 3D US volumes are translated into the customized intermediate modality using our ACMT approach. Based on the translated volumes, we employ our proposed anatomy-aware registration network to generate a warping map, which is subsequently applied to the original 3D MR to obtain the final warped 3D MR volume. In the following sections, we provide detailed explanations of these three core components of our framework.

\begin{figure}
    \centering
    \includegraphics[width=0.4\textwidth]{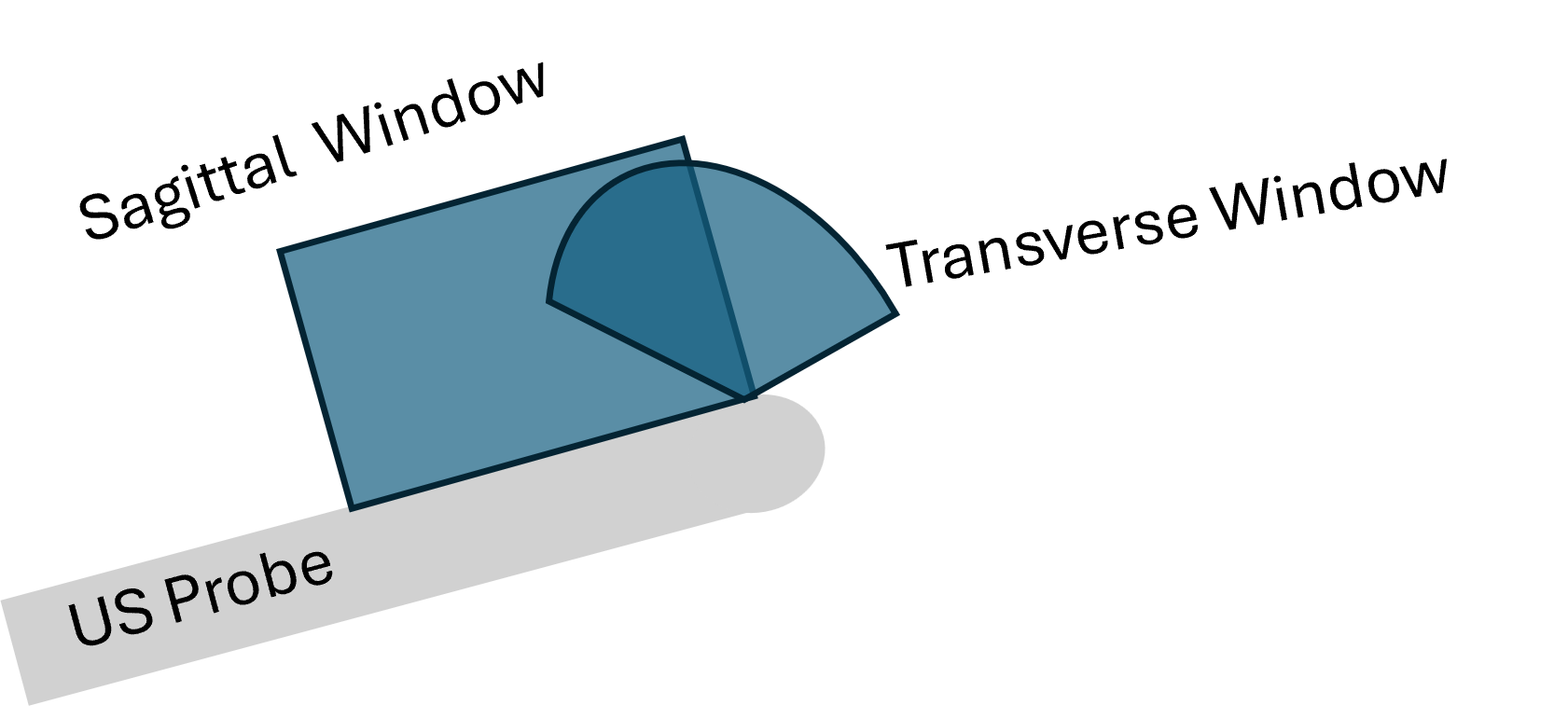}
    \caption{The bi-directional Scanning windows of TRUS}
    \label{fig:bi-directonal_scan}
\end{figure}

\begin{figure*}
    \centering
    \includegraphics[width=1.0\textwidth]{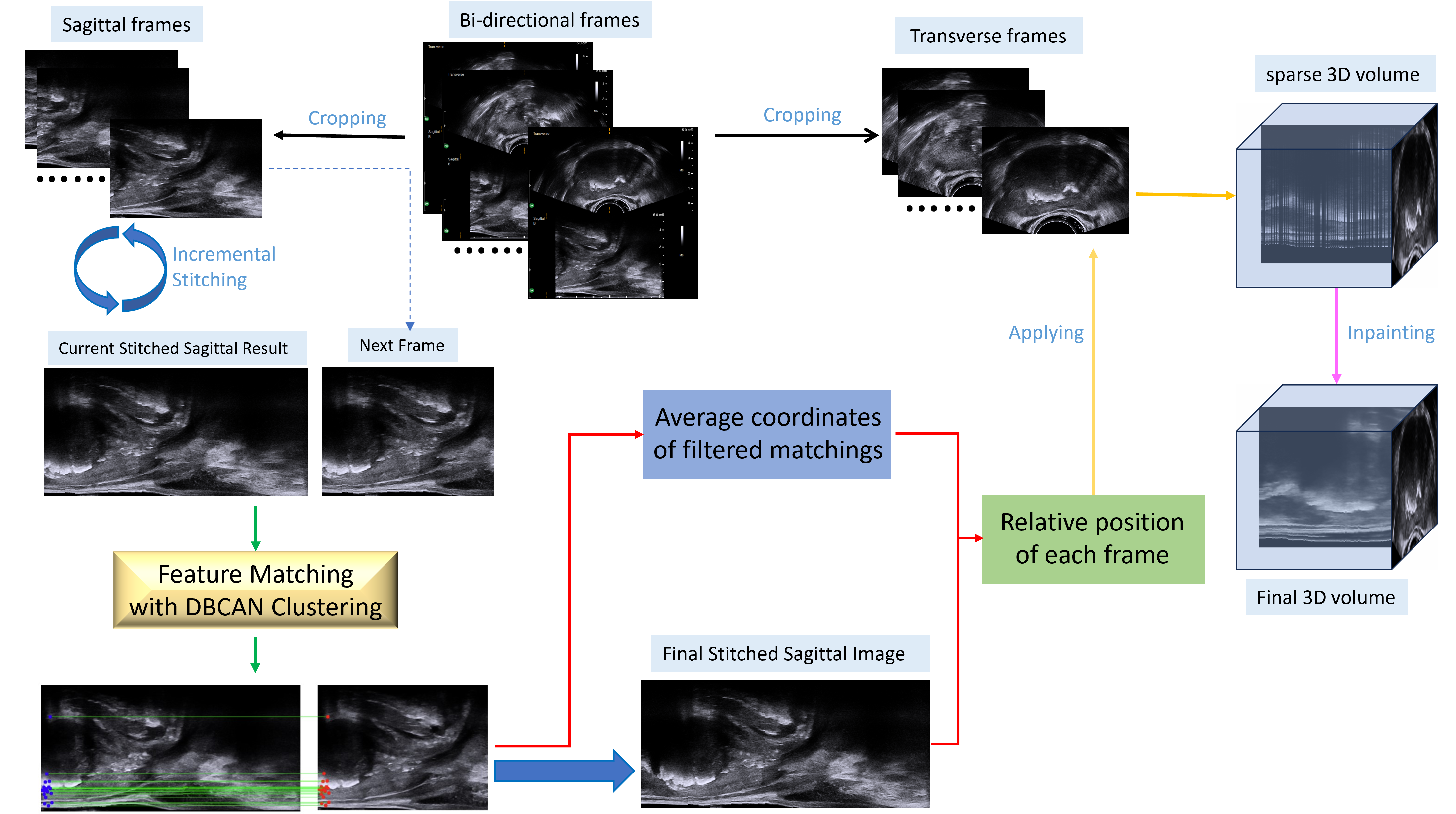}
    \caption{Our Probe-location-independent TRUS 3D reconstruction workflow}
    \label{fig:3D_reconstruction}
\end{figure*}
\subsection{Probe-Location-Independent TRUS 3D Reconstruction}
A key advantage of modern TRUS systems is their ability to simultaneously acquire orthogonal sagittal and transverse views. As shown in Fig. \ref{fig:bi-directonal_scan}, the probe acquires two ultrasound images simultaneously at each time point during biopsy: a rectangular sagittal frame aligned with the probe's long axis, and a fan-shaped transverse frame emitted from the probe tip. Some concrete examples of such bi-directional imaging can be seen from the Bi-directional frames in Fig. \ref{fig:3D_reconstruction}. These complementary perspectives facilitate 3D reconstruction without the need for external probe tracking. To capture the necessary data, only a brief initial scan is required, during which the physician keeps the probe rotationally fixed while moving it from anterior to posterior. This short acquisition phase enables complete 3D reconstruction while maintaining flexibility for unrestricted probe manipulation throughout the remainder of the procedure.

Since the probe is required to maintain a fixed orientation during the initial scan, the sagittal frames can be treated as overlapping patches from a single continuous sagittal plane. This enables full-plane reconstruction via image stitching. As illustrated by the blue thick arrows in Fig.~\ref{fig:3D_reconstruction}, our goal is to iteratively stitch the current stitched sagittal result with the next frame, via some feature matching methods, to progressively construct the final stitched sagittal image. However, the high noise level in ultrasound images severely affects the reliability of conventional feature matching methods (e.g., SIFT~\cite{rublee2011orb}, ORB~\cite{rublee2011orb}) and even advanced deep models (e.g., LoFTR~\cite{sun2021loftr}). This challenge is further amplified in our setting, where each patient’s ultrasound video contains hundreds of frames. This means that any error introduced during the stitching of a single frame will propagate and accumulate throughout the stitching of all subsequent frames. This raises higher demands on the robustness and reliability of the feature matching algorithms.

\begin{figure*}
    \centering
    \includegraphics[width=1.0\textwidth]{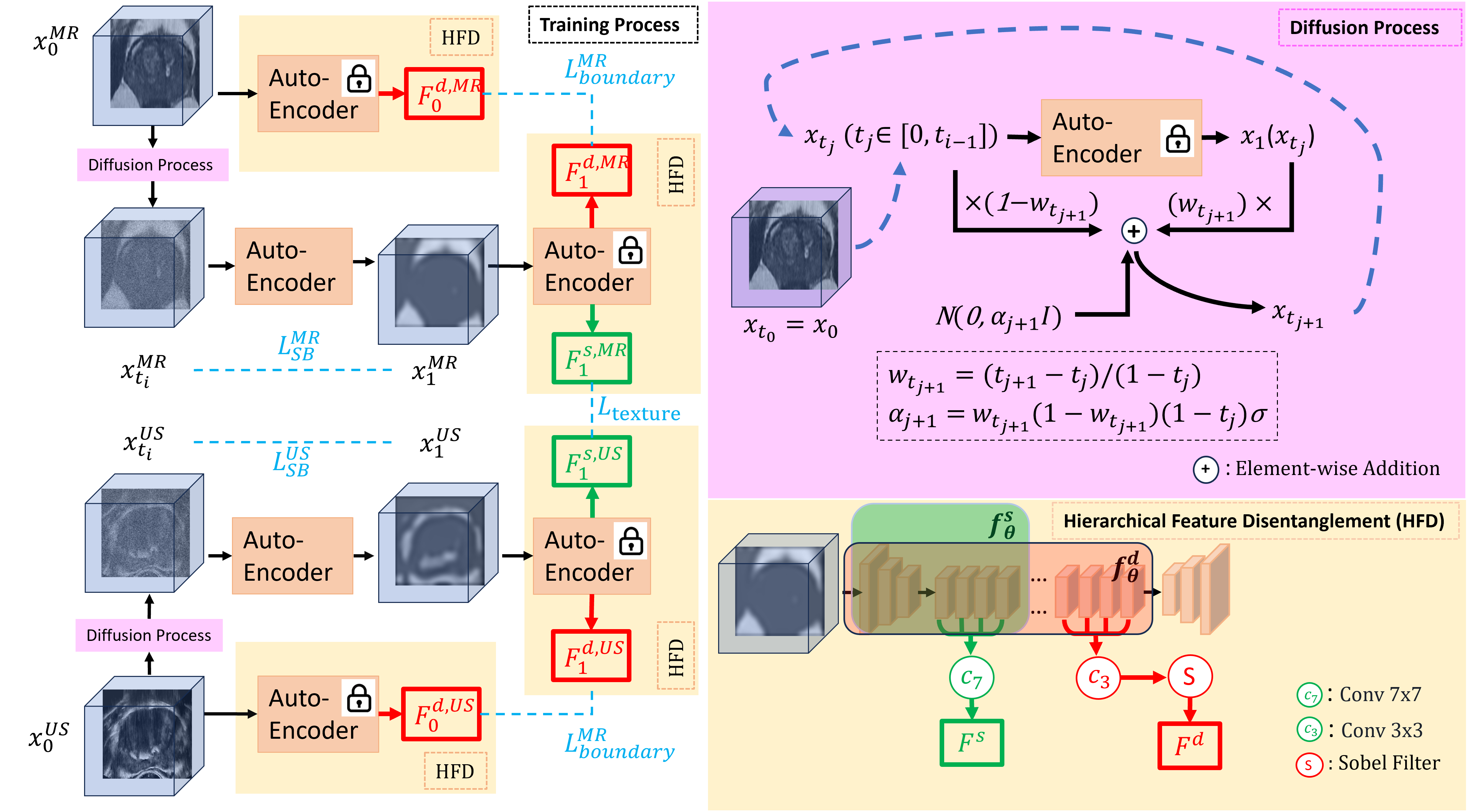}
    \caption{The Anatomically Coherent Modality Translation Framework}
    \label{fig:modality_translation_network}
\end{figure*}

To address this issue, as shown in the yellow box of Fig.~\ref{fig:3D_reconstruction}, we introduce a clustering-based filtering strategy to refine the matching points extracted by the feature matching methods. We leverage the fact that, in our case, the primary transformation between adjacent frames is simple translation, with negligible rotation or scale changes. This means that correctly matched points should exhibit consistent coordinate differences. Therefore, we apply DBSCAN ~\cite{deng2020dbscan} to cluster the coordinate differences of all matches, and retain only the largest cluster as inliers, effectively removing outliers. We choose DBSCAN here beacuse it forms clusters by identifying regions with sufficiently high data density while labeling points in low-density regions as noise. Unlike k-means, DBSCAN does not require specifying the number of clusters in advance. This makes it more suitable for our problem where the number of outlier categories is unknown. Thus, our method achieves robust and smooth final stitching result. For each patient, we evaluated SIFT+clustering, ORB+clustering, and LoFTR+clustering, and selected the best-performing approach.

As illustrated by the red arrows in Fig.~\ref{fig:3D_reconstruction}, while incrementally stitching each new sagittal frame to the current composite image, we also record the average coordinates of the filtered matching points between each incoming frame and the current stitched result. These coordinates are used to infer the relative position of each frame within the final stitched sagittal image.

After that, as shown by the yellow arrows in Fig.~\ref{fig:3D_reconstruction}, given the orthogonal relationship between sagittal and transverse views, the relative position information of the sagittal frames can be directly transferred to the corresponding transverse frames, determining their spatial positions along the third dimension. This enables the generation of the initial sparse 3D volume.

Finally, to complete the 3D reconstruction, we perform missing voxel inpainting on the sparse volume using Deep Image Prior (DIP) \cite{ulyanov2018deep}, as shown by the purple arrow in Fig.~\ref{fig:3D_reconstruction}. Leveraging the intrinsic feature-learning capability of convolutional networks, DIP fills in the missing regions through an end-to-end auto-encoder structure, producing a high-resolution and visually coherent 3D result.

Notably, another key innovation in our approach is the decision to perform stitching before frame localization, instead of directly comparing consecutive frames to obtain their relative positions. This approach offers several advantages. First, stitching creates a unified reference image with a broader field of view, making it easier to find more robust matching points. Additionally, it ensures consistency throughout the entire sequence. If only two consecutive frames are compared at a time, any matching error will lead to incorrect spatial localization of the current frame, which will then cause all subsequent frames to be mislocalized, as each frame's position depends solely on its matching with the previous one. In contrast, by matching each new frame to an already stitched result, even if an error occurs in a single frame, it does not necessarily cause mislocalization of the other frames. Furthermore, the smoothness of the final stitched image also directly reflects the quality of the 3D reconstruction. If there are any significant misalignments during stitching, we can visually detect discontinuities in the stitched image, allowing us to adjust the parameters before proceeding to inpainting. This provides an early check on the accuracy of the stitching, ensuring that the subsequent steps are more reliable.

\subsection{Anatomically Coherent Modality Translation}
After addressing the dimensional mismatch between MR and US images, we turn to bridging the modality gap. Our proposed ACMT framework, illustrated in Fig. \ref{fig:modality_translation_network}, is built on a Schrödinger Bridge-based diffusion model tailored to enhance anatomical consistency across modalities. The following section introduces the Schrödinger Bridge concept in Diffusion Models and detail our translation framework.

\subsubsection{Diffusion Schrödinger Bridge}
The standard diffusion model is designed to evolve from pure noise to a specific target distribution. To extend its capability for translation between arbitrary distributions, the Schrödinger Bridge (SB) framework was introduced in~\cite{kim2024unpaired}. Inspired by optimal transport theory \cite{wang2021deep,leonard2014survey}, SB formulates the problem as finding the optimal stochastic process that evolves a source distribution \( P_0 \) into a target distribution \( P_1 \) through a sequence of intermediate distributions \( P_t \) over time \( t \). Mathematically, this process is defined as:
\begin{equation}
P^{SB} = \{\argmin_{P_t} D_{KL}(P_t \lVert W^\sigma)\} \quad  with\quad t\sim[0,1],
\label{eq:SB_formula}
\end{equation}
where \( W^\sigma \) represents the Wiener measure with variance \( \sigma \). This formulation seeks to minimize the Kullback-Leibler (KL) divergence $D_{KL}$ between the process distribution \( P_t \) and the reference measure \( W^\sigma \) at each timestep $t$. The set of optimal distributions \( \{P_t^{SB}\} \) forms the Schrödinger Bridge \( P^{SB} \) , establishing a stochastic transformation between \( P_0^{SB} \) and \( P_1^{SB} \).

In practical implementations, we utilize a discretized approximation of this continuous process to ensure computational feasibility while maintaining its theoretical principles. Among the various solution strategies, the Conditional Flow Matching (CFM) formulation \cite{kim2024unpaired,tong2023conditional} has demonstrated exceptional effectiveness. CFM establishes that for any two distributions \( P_{t_m}^{SB} \) and \( P_{t_n}^{SB} \) within the Schrödinger Bridge, where \( [t_m,t_n] \subseteq [0,1] \), the intermediate distribution at time \( t \in [t_m,t_n] \) follows a Gaussian distribution:

\begin{multline}
P(X_t|X_{t_m},X_{t_n}) = \mathcal{N}\Big(X_t \Big| w_tX_{t_n}+(1-w_t)X_{t_m}, \\
w_t(1-w_t)\sigma(t_n-t_m)\textbf{\textit{I}}\Big),
\label{eq:conditional}
\end{multline}
where \( X_t \sim P_t^{SB} \), \( X_{t_m} \sim P_{t_m}^{SB} \), \( X_{t_n} \sim P_{t_n}^{SB} \), and \( w_t = {(t - t_m)}/{(t_n - t_m)} \).  

Furthermore, the joint distribution \( P_{t_m t_n}^{SB} \) between any two timesteps can be obtained by solving an entropy-regularized optimal transport problem, formulated as follows:  

\begin{multline}
P_{t_m t_n}^{SB} = \argmin_{P_{t_m,t_n}} \mathbb{E}_{(X_{t_m},X_{t_n})}[\lVert X_{t_m} - X_{t_n} \rVert^2] \\
- 2\sigma (t_n - t_m) H(X_{t_m}, X_{t_n}),
\label{eq:optjoint}
\end{multline}  
where \( H \) represents the entropy function. This formulation allows for the optimal determination of the terminal distribution \( P_{t_n}^{SB} \) when given an initial distribution \( P_{t_m}^{SB} \). By applying Equation (\ref{eq:conditional}), we can then compute any intermediate distribution \( P_t^{SB} \) along the Schrödinger Bridge, facilitating a smooth transition between the initial and target distributions.

\subsubsection{Modality Translation Workflow}
Building on the Schrödinger Bridge-based diffusion model, we propose an anatomically coherent modality translation framework. We assume that both MR and US images can be mapped to a shared intermediate modality that preserves only boundary information, one of the few features consistently visible across both modalities, while suppressing modality-specific tissue details. To achieve this, we construct a Schrödinger Bridge from either the MR distribution \( P_0^{MR} \) or the US distribution \( P_0^{US} \) 
  to the intermediate modality \( P_1 \). By filtering out non-essential textures and internal differences, this transformation effectively reduces modality discrepancies, providing a more reliable basis for registration.

Specifically, our modality translation workflow consists of two phases: the diffusion process and the training process.

The diffusion process is illustrated in the purple block of Fig. ~\ref{fig:modality_translation_network}. We define \( x_{t_j} \in P_{t_j} \) as an intermediate state along the Schrödinger Bridge at time \( t_j \in [0,1] \). The goal of our network \( f_\theta \) is to map \( x_{t_j} \) to the terminal state \( x_1 \in P_1 \). Once \( x_1 \) is obtained, the next state \( x_{t_{j+1}} \) can be derived using the Conditional Flow Matching (CFM) formulation in Equation~\ref{eq:conditional}:
\begin{equation}
x_{t_{j+1}} = w_{t_{j+1}} x_1 + (1 - w_{t_{j+1}}) x_{t_j} + \mathcal{N}(0, \alpha_{j+1} I),
\label{eq:diffusion}
\end{equation}
where \( w_{t_{j+1}} = {(t_{j+1} - t_j)}/{(1 - t_j)} \) represents the interpolation weight between \( x_1 \) and \( x_{t_j} \). The Gaussian noise term \( \mathcal{N}(0, \alpha_{j+1} I) \) is scaled by $\alpha_{j+1}$. It controls the noise magnitude via $\alpha_{j+1} = w_{t_{j+1}} (1 - w_{t_{j+1}}) (1 - t_j) \sigma$, where $\sigma$ represents the variance in Equation \ref{eq:SB_formula}. This iterative process begins with \(x_{t_0} = x_0\), the source MR or US image, and progressively transforms it towards \(x_{t_i}\).

Upon the clarity of the diffusion process, the model training will proceed. As illustrated in the Training Process (left side of Fig.~\ref{fig:modality_translation_network}), during this phase, we follow the steps outlined in Algorithm~\ref{alg:modality_translation} for each MR-US image pair.
    

\begin{algorithm}
\caption{Training Process for Modality Translation}
\label{alg:modality_translation}
\begin{algorithmic}[1]
\STATE Randomly selected $t_i$ from the predefined sample pool [$t_0$, $t_1$,$t_2$,...,$t_T$] where any $t_i \sim [0,1]$

\STATE \textbf{Switch the model $f_\theta$ to evaluation mode} (with all parameters locked)
\STATE Iteratively generate $x_{t_i}$ from $x_0$ for both MR and US via the diffusion process as outlined in the purple block of Fig. \ref{fig:modality_translation_network}.

\STATE \textbf{Switch the model $f_\theta$ to training mode}
\STATE Compute transformation from $x_{t_i}$ to $x_1$ for both MR and US: $x_1 \leftarrow f_\theta(x_{t_i})$
\STATE Compute loss $\mathcal{L}$
\STATE Update parameters $\theta$ via backpropagation using $\mathcal{L}$
\end{algorithmic}
\end{algorithm}

Through this training strategy, the network learns to map any transitional sample \( x_{t_i} \) to the target modality \( x_1 \), effectively capturing the full transformation along the Schr\"{o}dinger Bridge. 


\subsubsection{Loss Functions}
First of all, our framework incorporates a SB constraint loss to ensure the transformation follows the optimal transport path defined by the Schr\"{o}dinger Bridge. This loss is derived from the joint distribution \( P_{t_i,1}^{SB} \), as defined in Equation~\ref{eq:optjoint}. For MR and US images, the SB constraint losses are defined as follows:

\begin{equation}
\begin{aligned}
\mathcal{L}_{\text{SB}}^{MR}(\theta_i,t_i) &= \mathbb{E}_{(x_{t_i}^{MR},x_1^{MR})} \left[ \| x_{t_i}^{MR} - x_1^{MR} \|^2 \right] \\
&\quad - 2\sigma(1-t_i)H(x_{t_i}^{MR},x_1^{MR}),
\end{aligned}
\end{equation}
\begin{equation}
\begin{aligned}
\mathcal{L}_{\text{SB}}^{US}(\theta_i,t_i) &= \mathbb{E}_{(x_{t_i}^{US},x_1^{US})} \left[ \| x_{t_i}^{US} - x_1^{US} \|^2 \right] \\
&\quad - 2\sigma(1-t_i)H(x_{t_i}^{US},x_1^{US}).
\end{aligned}
\end{equation}
where \( x_1^{MR} = f_{\theta_i}(x_{t_i}^{MR}) \) and \( x_1^{US} = f_{\theta_i}(x_{t_i}^{US}) \) represent the terminal states predicted by the network. The total SB constraint loss is then defined as:
\begin{equation}
\mathcal{L}_{\text{SB}} =  \left( \mathcal{L}_{\text{SB}}^{MR} + \mathcal{L}_{\text{SB}}^{US} \right)/2.
\end{equation}

However, the Schr\"{o}dinger Bridge loss alone cannot explicitly guide the generation of anatomically meaningful features. To address this, we introduce hierarchical anatomy consistency losses: a deep-layer boundary loss and a shallow-layer texture loss. Deeper layers of the network are more effective at capturing fine-grained, structured information such as anatomical boundaries, while shallower layers are better suited for modeling general, less structured features like textures and internal prostate regions. Accordingly, the boundary loss encourages accurate preservation of prostate contours, and the texture loss promotes consistent internal representations by smoothing out modality-specific variations. 

As illustrated in the \textit{Hierarchical Feature Disentanglement} block of Fig.~\ref{fig:modality_translation_network}, let \(f_\theta^s\) and \(f_\theta^d\) represent the shallow and deep feature extraction functions of our network, respectively. Given an input image \(x_0\) (either MR or US), the shallow and deep features are extracted as follows:
\begin{equation}
\mathbf{F}_0^s = f_\theta^s(x_0), \quad \mathbf{F}_0^d = f_\theta^d(x_0).
\end{equation}
Similarly, for the transformed image \(x_1\) in the intermediate modality \(P_1\), the corresponding features are extracted as:
\begin{equation}
\mathbf{F}_1^s = f_\theta^s(x_1), \quad \mathbf{F}_1^d = f_\theta^d(x_1).
\end{equation}
On one hand, to preserve the anatomical boundaries of the original images, we process the deep features \(\mathbf{F}_0^d\) and \(\mathbf{F}_1^d\) using a smaller convolutional kernel (\(3 \times 3\)), followed by a Sobel filter. This design choice is motivated by the fact that smaller kernels are particularly effective at extracting fine-grained features, such as edges and boundaries, as they focus on localized regions and capture high-frequency details. Therefore, the boundary preservation loss for MR and US images is defined as:
\begin{equation}
\mathcal{L}_{\text{boundary}}^{MR} = \left\| \mathcal{S}(\mathcal{C}_{3\times3}(\mathbf{F}_1^{d,MR})) - \mathcal{S}(\mathcal{C}_{3\times3}(\mathbf{F}_0^{d,MR})) \right\|_1,
\end{equation}
\begin{equation}
\mathcal{L}_{\text{boundary}}^{US} = \left\| \mathcal{S}(\mathcal{C}_{3\times3}(\mathbf{F}_1^{d,US})) - \mathcal{S}(\mathcal{C}_{3\times3}(\mathbf{F}_0^{d,US})) \right\|_1,
\end{equation}
where \(\mathcal{C}_{3\times3}(\cdot)\) denotes the \(3 \times 3\) convolution operation, \(\mathcal{S}(\cdot)\) represents the Sobel filter, and \(\|\cdot\|_1\) denotes the L1 norm. The total boundary preservation loss is given by:
\begin{equation}
\mathcal{L}_{\text{boundary}} =  \left( \mathcal{L}_{\text{boundary}}^{MR} + \mathcal{L}_{\text{boundary}}^{US} \right)/2.
\end{equation}

On the other hand, to ensure texture and prostate internal coherence between the translated MR and US images in the intermediate domain, we further process the shallow features \(\mathbf{F}_1^{s,MR}\) and \(\mathbf{F}_1^{s,US}\) using a larger convolutional kernel (\(7 \times 7\)), as larger kernels are known to be more effective at capturing global features such as texture patterns due to their broader receptive fields. Hence, the texture consistency loss is formulated as:
\begin{equation}
\mathcal{L}_{\text{texture}} = \left\| \mathcal{C}_{7\times7}(\mathbf{F}_1^{s,MR}) - \mathcal{C}_{7\times7}(\mathbf{F}_1^{s,US}) \right\|_2^2,
\end{equation}
where \(\mathcal{C}_{7\times7}(\cdot)\) represents the \(7 \times 7\) convolution operation, and \(\|\cdot\|_2^2\) denotes the squared L2 norm.

Finally, the overall loss function is a combination of the SB constraint loss, boundary preservation loss and texture consistency loss:
\begin{equation}
\mathcal{L}_{\text{total}} = 
\lambda_{\text{SB}} \mathcal{L}_{\text{SB}} +
\lambda_{\text{boundary}} \mathcal{L}_{\text{boundary}} +
\lambda_{\text{texture}} \mathcal{L}_{\text{texture}}, 
\end{equation}
where the weighting coefficients \(\lambda_{\text{SB}}\), \(\lambda_{\text{boundary}}\), and \(\lambda_{\text{texture}}\) are carefully tuned to balance the contributions of each loss component, ensuring that the network simultaneously achieves optimal transport, boundary preservation and texture consistency.


\begin{figure*}
    \centering
    \includegraphics[width=0.8\textwidth]{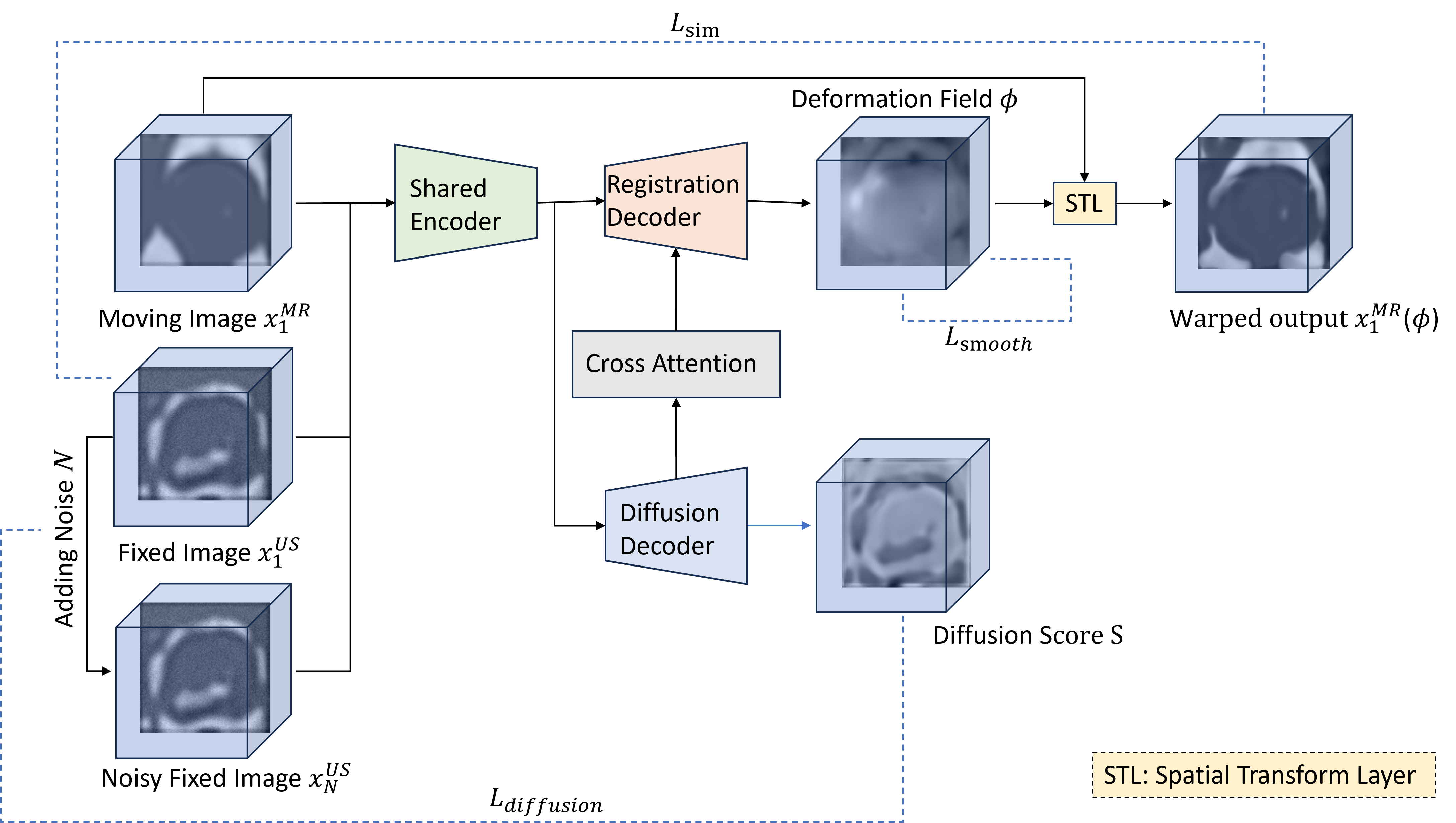}
    \caption{Anatomy-Aware Image Registration Network}
    \label{fig:regis_net}
\end{figure*}

\subsection{Anatomy-Aware Image Registration}


After employing our ACMT, the transformed MR and US images share similar internal prostate structures, while preserving distinct boundary shapes and thicknesses. To further ensure that the focus of registration is placed on the anatomically unified internal regions of the prostate, we propose an anatomy-aware diffusion-based registration model. We reformulate the registration objective to prioritize the alignment of the dark interior regions in the transformed images, where cross-modal consistency is maximized. 

\subsubsection{The Registration Workflow}
As shown in Fig. \ref{fig:regis_net}, the network takes as inputs the moving image \( x_1^{MR} \), the fixed image \( x_1^{US} \), and a noise-perturbed version of the fixed image \( x_N^{US} \). These are processed by a shared encoder into a common latent space. The encoded representations are then passed to two decoders: the registration decoder predicts the deformation field \(\Phi\), while the diffusion decoder generates the diffusion score \( S \). To enhance feature extraction, we incorporate the cross-attention mechanism from FSDiffReg~\cite{qin2023fsdiffreg}, enabling multi-scale guidance from the diffusion decoder to refine the registration decoder. Subsequently, the predicted deformation field \(\Phi\) is applied to the moving image \( x_1^{MR} \) using a spatial transform layer, yielding the warped image \( x_1^{MR}(\Phi) \).

Importantly, during inference, the deformation field is estimated from the translated images but applied to the original, unaltered inputs. This strategy reduces cross-modal inconsistencies during registration while preserving the full anatomical and intensity information of the original data. Consequently, the warped outputs remain faithful to the native modality, making them suitable for downstream tasks such as anatomical comparison, image fusion, and clinical interpretation.

\subsubsection{Loss Functions} 
To prioritize the alignment of the dark interior regions of the prostate while minimizing the impact of the high-information boundary areas, we introduce an anatomy-aware function that redefines the key regions of interest for registration.
\begin{equation}
F_{\text{Ana}}(x) = \text{sigmoid}(-(x - \text{mean}(x)))
\label{eq:anatomy-aware}
\end{equation}

By subtracting the mean intensity and applying a negative sign, the function adaptively normalizes the image, shifting low-intensity regions (prostate interior) to positive values and high-intensity regions (boundaries and bright structures) to negative values. A subsequent sigmoid transformation maps darker regions to values closer to 1, emphasizing their contribution to the loss, while brighter regions are mapped to values closer to 0, reducing their influence. This smooth, differentiable sigmoid transformation prevents abrupt thresholding, stabilizes optimization, and ensures anatomically meaningful alignment. The anatomy-aware similarity loss is then formulated as:

\begin{equation}
\mathcal{L}_{\text{sim}} = 1 - \frac{2\sum (F_{\text{Ana}}(x_1^{\text{US}}) F_{\text{Ana}}(x_1^{\text{MR}}(\Phi)))+\epsilon}{\sum \left( F_{\text{Ana}}(x_1^{\text{US}})\right) + \sum \left( F_{\text{Ana}}(x_1^{\text{MR}}(\Phi))\right) +\epsilon}
\end{equation}

We adopt a soft Dice loss to evaluate image similarity here. A small constant $\epsilon$ (set to $1 \times 10^{-6}$ in our experiments) is added to prevent division by zero and to ensure numerical stability. Unlike the traditional Dice loss, which is typically applied to binary masks, the soft Dice loss operates directly on continuous-valued inputs. It is both differentiable and well-suited to our registration task. The formulation naturally emphasizes the overlap of regions with higher values rather than treating all voxels equally, since higher-valued regions contribute more to reducing the loss. By the application of our anatomy-aware function $F_{\text{Ana}}$ in Eq.~\ref{eq:anatomy-aware}, we reweight voxel-wise contributions to highlight semantically consistent prostate interior regions while down-weighting the ambiguous boundary areas. As a result, the proposed loss encourages the network to focus more on anatomical alignment within the prostate so as to achieve anatomy-aware registration.

In addition, to further ensure the smoothness of the deformation, we apply a smoothness constraint on the deformation field by limiting its gradients. 
\begin{equation}
\mathcal{L}_{\text{smooth}} = \sum \| \nabla \Phi\|_2^2
\end{equation}

Finally, to ensure that the decoder outputs remain consistent with the underlying diffusion process, we incorporate a diffusion loss as the final objective. This loss enforces alignment between the predicted diffusion scores and the expected denoising purpose, thereby reinforcing the probabilistic structure learned through the diffusion model and enhancing the overall stability and fidelity of the registration framework.
\begin{equation}
\mathcal{L}_{\text{diffusion}} = \| S - N \|_2^2
\end{equation}

As illustrated in Figure~\ref{fig:regis_net}, $N$ corresponds to the Gaussian noise added to the fixed image prior to network input, while $S$ represents the score predicted by the diffusion decoder. 

To sum up, our complete anatomy-aware registration network combines three essential components:

\begin{equation}
\mathcal{L}_{\text{total}} = \lambda_{\text{sim}}\mathcal{L}_{\text{sim}} + \lambda_{\text{smooth}}\mathcal{L}_{\text{smooth}} + \lambda_{\text{diff}}\mathcal{L}_{\text{diffusion}}
\end{equation}

where the coefficients \( \lambda \) serve as trade-off parameters, balancing the contribution of each individual loss term within the overall objective function.

\begin{figure*}
    \centering

    \begin{subfigure}{0.12\textwidth}
        \centering \scriptsize \textbf{US}
    \end{subfigure}
    \begin{subfigure}{0.12\textwidth}
        \centering \scriptsize \textbf{MR}
    \end{subfigure}
    \begin{subfigure}{0.12\textwidth}
        \centering \scriptsize \textbf{UNSB-MR}
    \end{subfigure}  
    \begin{subfigure}{0.12\textwidth}
        \centering \scriptsize \textbf{PMT-US}
    \end{subfigure}
    \begin{subfigure}{0.12\textwidth}
        \centering \scriptsize \textbf{PMT-MR}
    \end{subfigure}
    \begin{subfigure}{0.12\textwidth}
        \centering \scriptsize \textbf{ACMT-US}
    \end{subfigure}
    \begin{subfigure}{0.12\textwidth}
        \centering \scriptsize \textbf{ACMT-MR}
    \end{subfigure}
    
    \vspace{0.5em} 
    
    \begin{subfigure}{0.12\textwidth}
        \centering
        \includegraphics[width=\textwidth]{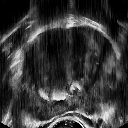}
    \end{subfigure}
    \begin{subfigure}{0.12\textwidth}
        \centering
        \includegraphics[width=\textwidth]{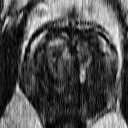}
    \end{subfigure}
    \begin{subfigure}{0.12\textwidth}
        \centering
        \includegraphics[width=\textwidth]{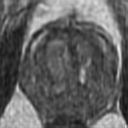}
    \end{subfigure}  
    \begin{subfigure}{0.12\textwidth}
        \centering
        \includegraphics[width=\textwidth]{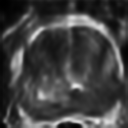}
    \end{subfigure}
    \begin{subfigure}{0.12\textwidth}
        \centering
        \includegraphics[width=\textwidth]{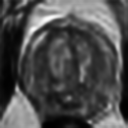}
    \end{subfigure}
    \begin{subfigure}{0.12\textwidth}
        \centering
        \includegraphics[width=\textwidth]{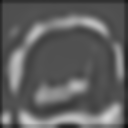}
    \end{subfigure}
    \begin{subfigure}{0.12\textwidth}
        \centering
        \includegraphics[width=\textwidth]{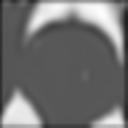}
    \end{subfigure}
    
    \vspace{0.1em} 
    
    \begin{subfigure}{0.12\textwidth}
        \centering
        \includegraphics[width=\textwidth]{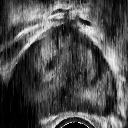}
    \end{subfigure}
    \begin{subfigure}{0.12\textwidth}
        \centering
        \includegraphics[width=\textwidth]{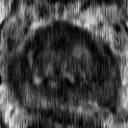}
    \end{subfigure}
    \begin{subfigure}{0.12\textwidth}
        \centering
        \includegraphics[width=\textwidth]{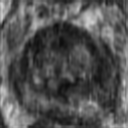}
    \end{subfigure}  
    \begin{subfigure}{0.12\textwidth}
        \centering
        \includegraphics[width=\textwidth]{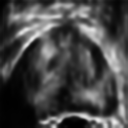}
    \end{subfigure}
    \begin{subfigure}{0.12\textwidth}
        \centering
        \includegraphics[width=\textwidth]{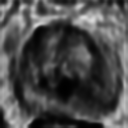}
    \end{subfigure}
    \begin{subfigure}{0.12\textwidth}
        \centering
        \includegraphics[width=\textwidth]{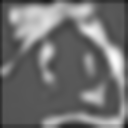}
    \end{subfigure}
    \begin{subfigure}{0.12\textwidth}
        \centering
        \includegraphics[width=\textwidth]{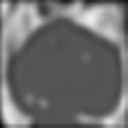}
    \end{subfigure}

    \caption{Visual comparison of modality translation results for two patients: Each row corresponds to one patient, showing the original US, original MR, MR-to-US translation (UNSB), and intermediate modality images generated by PMT and our ACMT.}
    \label{fig:translation}
\end{figure*}

\section{Results and Evaluation}
\subsection{Dataset}
The study utilized 5 pairs of T2-weighted volumetric MR scans and 2D US biopsy videos collected from Southmead Hospital Bristol. The MR data are inherently 3D, with anisotropic resolution due to larger inter-slice spacing. To enhance anatomical continuity, we applied deep image prior \cite{ulyanov2018deep} to perform inpainting. For US, which only contains 2D frames, we employed our proposed probe-location-independent 3D reconstruction method to generate volumetric data. All resulting volumes were cropped around the prostate while preserving key anatomical context for downstream tasks.

The data were split into training (80\%) and testing (20\%) sets, with cross-validation conducted in all experiments. Given the limited dataset size, we applied extensive data augmentation through random flipping and rotation during the training. To balance computational efficiency and anatomical preservation, all volumes were resampled to a standardized resolution of \(128 \times 128 \times 64\) voxels as the input of registration network.

\subsection{Evaluation}

We evaluate our method in two key areas. First, we demonstrate the superior ability of our modality translation method to reduce inter-modal discrepancies compared to state-of-the-art (SOTA) methods, including UNSB \cite{kim2024unpaired}, a modality-translation technique debuted at ICLR 2024, and our earlier method PMT \cite{ma2024pmt}. Second, we show that our complete framework, which combines anatomically coherent modality translation with anatomy-aware registration, outperforms existing SOTA registration methods like DiffusionMorph \cite{kim2022diffusemorph} and FSdiffReg \cite{qin2023fsdiffreg}. Additionally, we separately evaluate the contributions of both our modality translation and registration components, demonstrating that each part enhances performance.

\subsubsection{Modality Translation Performance}
To quantitatively evaluate the modality translation performance, we adopted two widely-used metrics: the Fréchet Inception Distance (FID) and Kernel Inception Distance (KID) \cite{chen2020reusing}. The FID measures the Wasserstein-2 distance between distributions using a pre-trained Inception-v3 network, where lower values indicate better distribution matching. The KID provides an unbiased alternative to FID that is particularly robust for smaller sample sizes. Both metrics comprehensively assess the quality of the modality translation. 

\begin{table}[t]
    \centering
    \caption{Quantitative evaluation of modality translation quality using FID and KID (lower is better).}
    \label{tab:fid_kid}
    \begin{tabularx}{\linewidth}{Xcc}  
        \toprule
        Method & FID $\downarrow$ (decrease by $\uparrow$) & KID $\downarrow$ (decrease by $\uparrow$)  \\
        \midrule
        Original & 404.88 & 0.56 \\
        UNSB & 377.92 (6.66\%) & 0.52 (7.14\%)  \\
        PMT  & 170.02 (58.01\%) & 0.11 (80.36\%) \\
        ACMT(Ours) & \textbf{138.01 (65.91\%)} & \textbf{0.09 (83.93\%)} \\
        \bottomrule
    \end{tabularx}
\end{table}

After cross-validation, the average results presented in Table~\ref{tab:fid_kid} demonstrate that our method outperforms existing approaches in terms of both FID and KID. Specifically, our method reduces FID by 65.91\% and KID by 83.93\%, outperforming UNSB by approximately 10-fold and 12-fold, respectively. Compared to our previously proposed PMT method, the current approach further improves both metrics, highlighting its enhanced capability in modality translation.

Visually, as shown in Figure \ref{fig:translation}, while UNSB performs unidirectional MR-to-US translation, it introduces US-specific artifacts like speckle noise and acoustic shadowing, which appear as granular white noise on the image surface. This superficial transformation, which merely mimics visual characteristics, offers limited value for image registration. PMT, although mapping both modalities into a shared intermediate space to align noise patterns and textures, still suffers from inconsistent anatomical representations of the prostate. The intensity variations in the prostate region carry different clinical meanings across modalities, which inevitably affects registration accuracy. In contrast, our approach is the first to simultaneously achieve boundary preservation and anatomical standardization. It maintains clear demarcation between internal and external prostate regions while maximizing the consistency of internal anatomical feature representations. This marks a significant advancement in medical image translation.

\begin{figure*}
    \centering

    \begin{subfigure}{0.117\textwidth}
        \centering
        \textbf{US}\\[0.3em]
        \includegraphics[width=\textwidth]{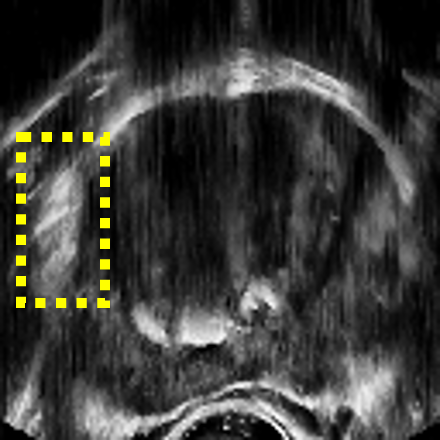}
    \end{subfigure}
    \begin{subfigure}{0.117\textwidth}
        \centering
        \textbf{MR}\\[0.3em]
        \includegraphics[width=\textwidth]{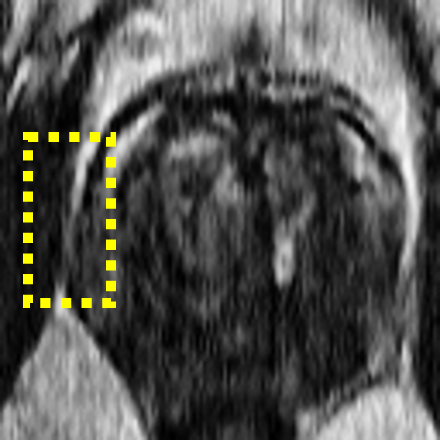}
    \end{subfigure}
    \begin{subfigure}{0.117\textwidth}
        \centering
        \textbf{DiffuseMorph}\\[0.3em]
        \includegraphics[width=\textwidth]{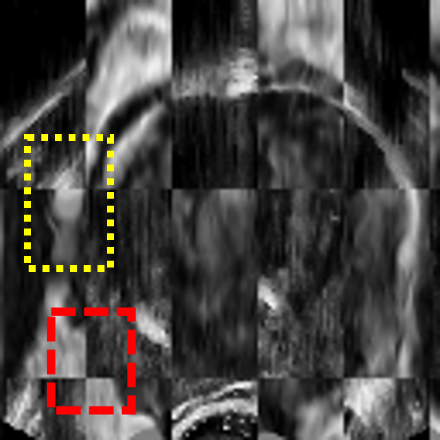}
    \end{subfigure}
    \begin{subfigure}{0.117\textwidth}
        \centering
        \textbf{FSDiffReg}\\[0.3em]
        \includegraphics[width=\textwidth]{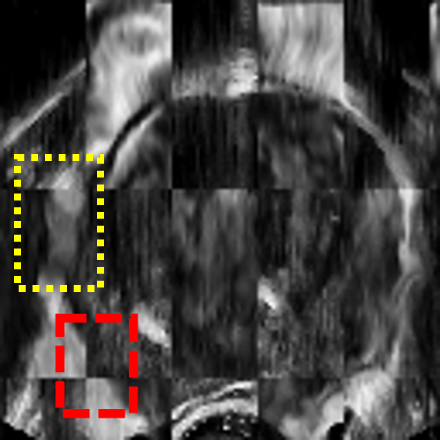}
    \end{subfigure}
    \begin{subfigure}{0.117\textwidth}
        \centering
        \textbf{UNSB+ FSDiffReg}\\[0.3em]
        \includegraphics[width=\textwidth]{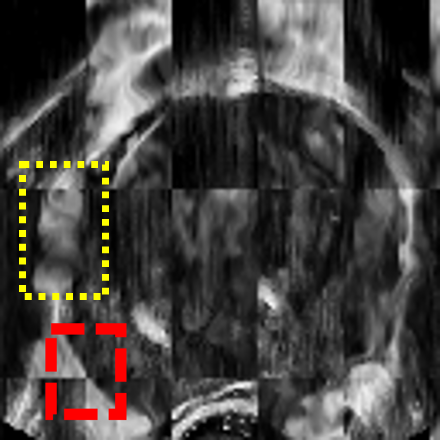}
    \end{subfigure}
    \begin{subfigure}{0.117\textwidth}
        \centering
        \textbf{PMT+ FSDiffReg}\\[0.3em]
        \includegraphics[width=\textwidth]{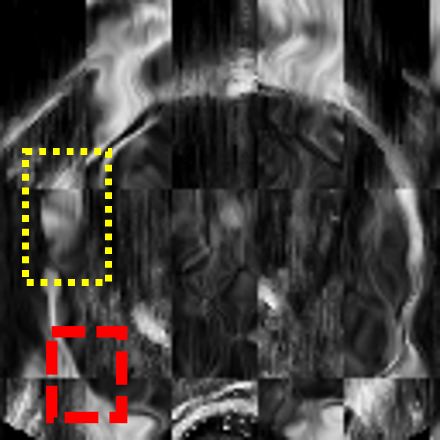}
    \end{subfigure}
    \begin{subfigure}{0.117\textwidth}
        \centering
        \textbf{ACMT+ FSDiffReg}\\[0.3em]
        \includegraphics[width=\textwidth]{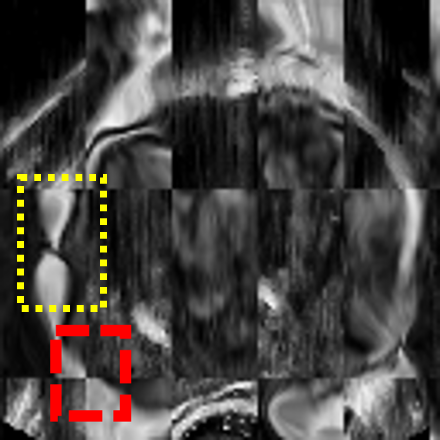}
    \end{subfigure}
    \begin{subfigure}{0.117\textwidth}
        \centering
        \textbf{MR2US-Pro (Ours)}\\[0.3em]
        \includegraphics[width=\textwidth]{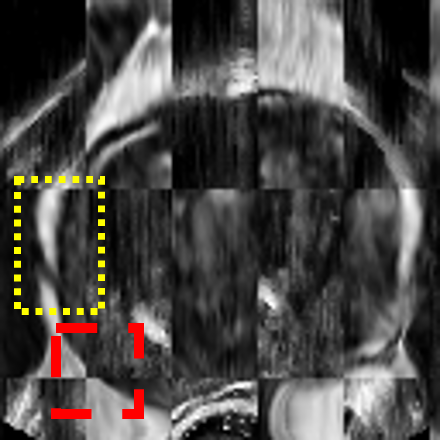}
    \end{subfigure}
    \\[0.1em] 
    \begin{subfigure}{0.117\textwidth}
        \centering
        \includegraphics[width=\textwidth]{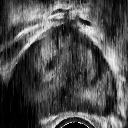}
    \end{subfigure}
    \begin{subfigure}{0.117\textwidth}
        \centering
        \includegraphics[width=\textwidth]{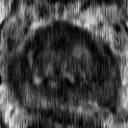}
    \end{subfigure}
    \begin{subfigure}{0.117\textwidth}
        \centering
        \includegraphics[width=\textwidth]{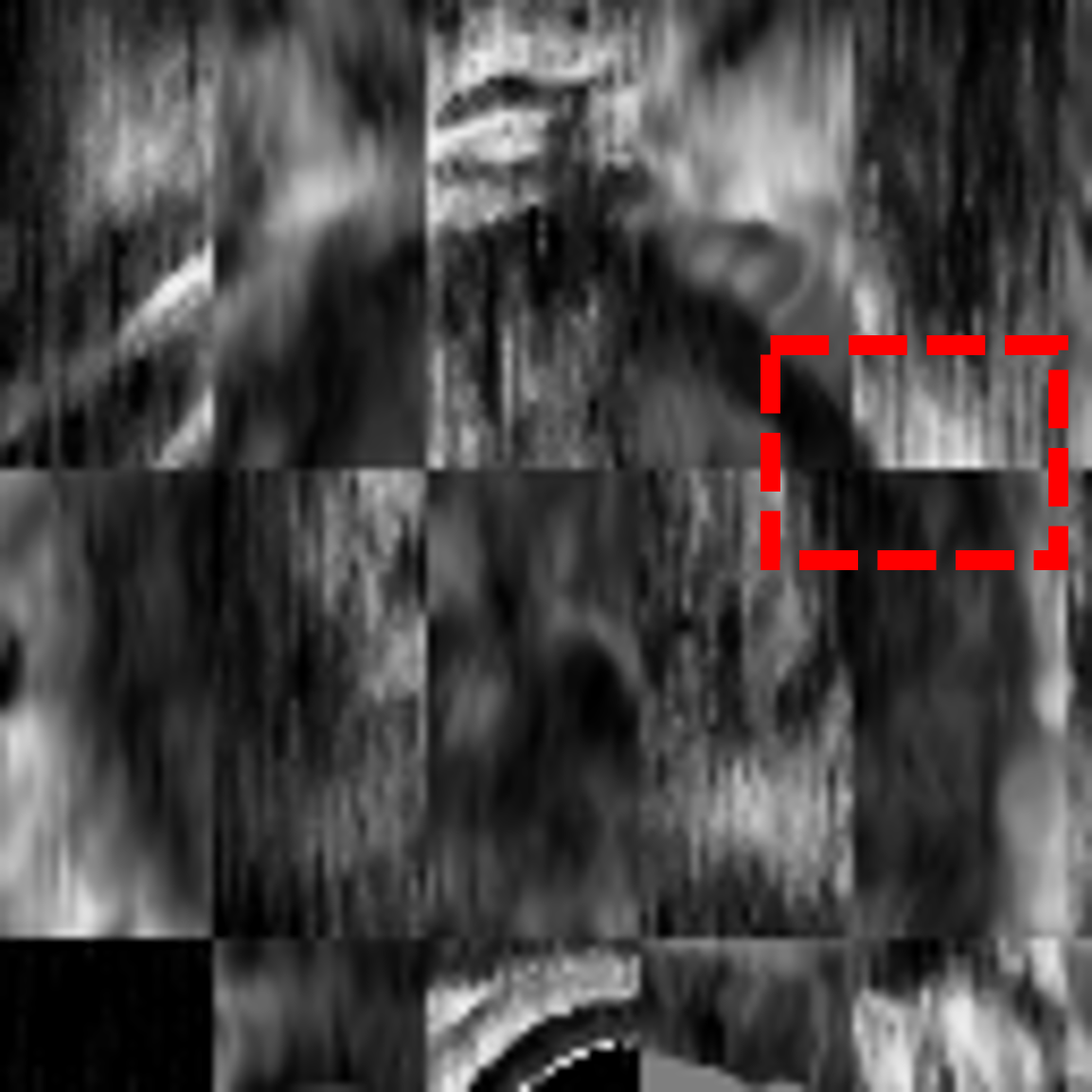}
    \end{subfigure}
    \begin{subfigure}{0.117\textwidth}
        \centering
        \includegraphics[width=\textwidth]{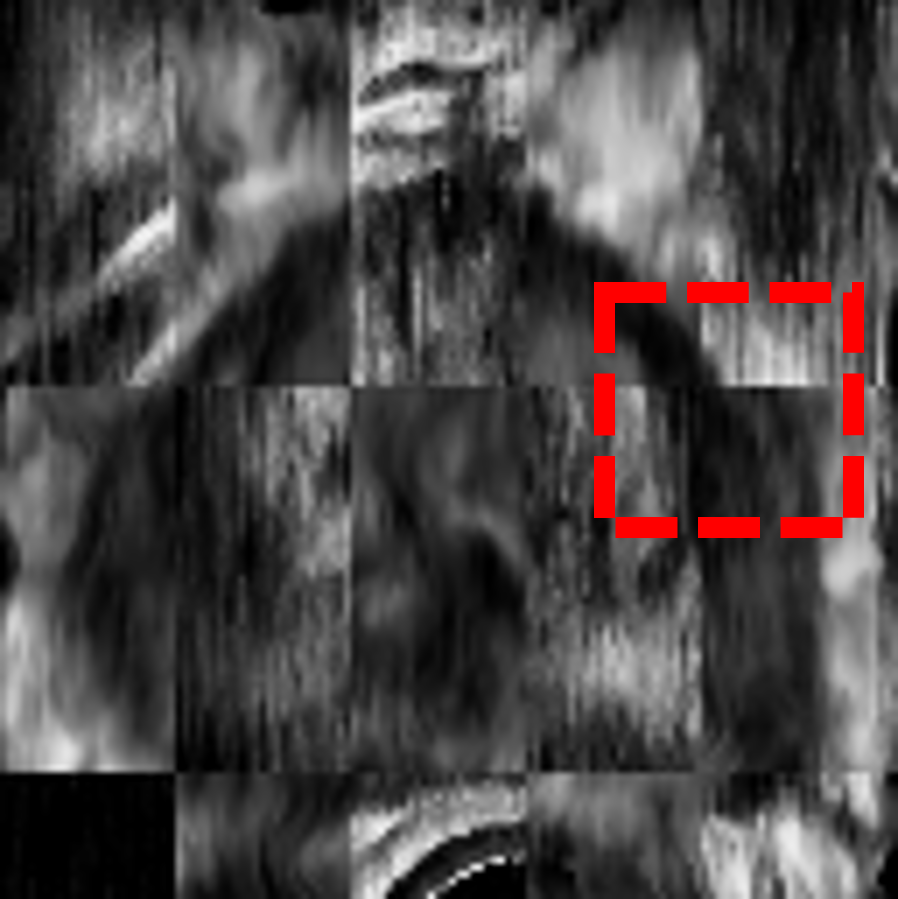}
    \end{subfigure}
    \begin{subfigure}{0.117\textwidth}
        \centering
        \includegraphics[width=\textwidth]{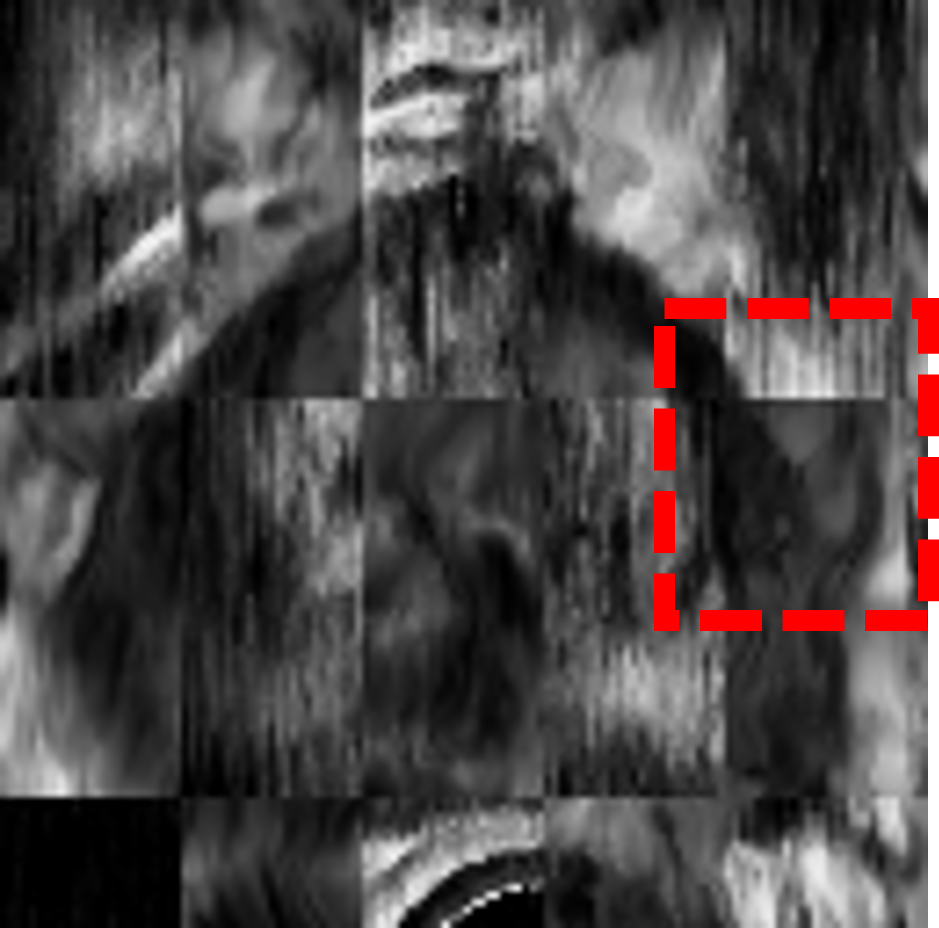}
    \end{subfigure}
    \begin{subfigure}{0.117\textwidth}
        \centering
        \includegraphics[width=\textwidth]{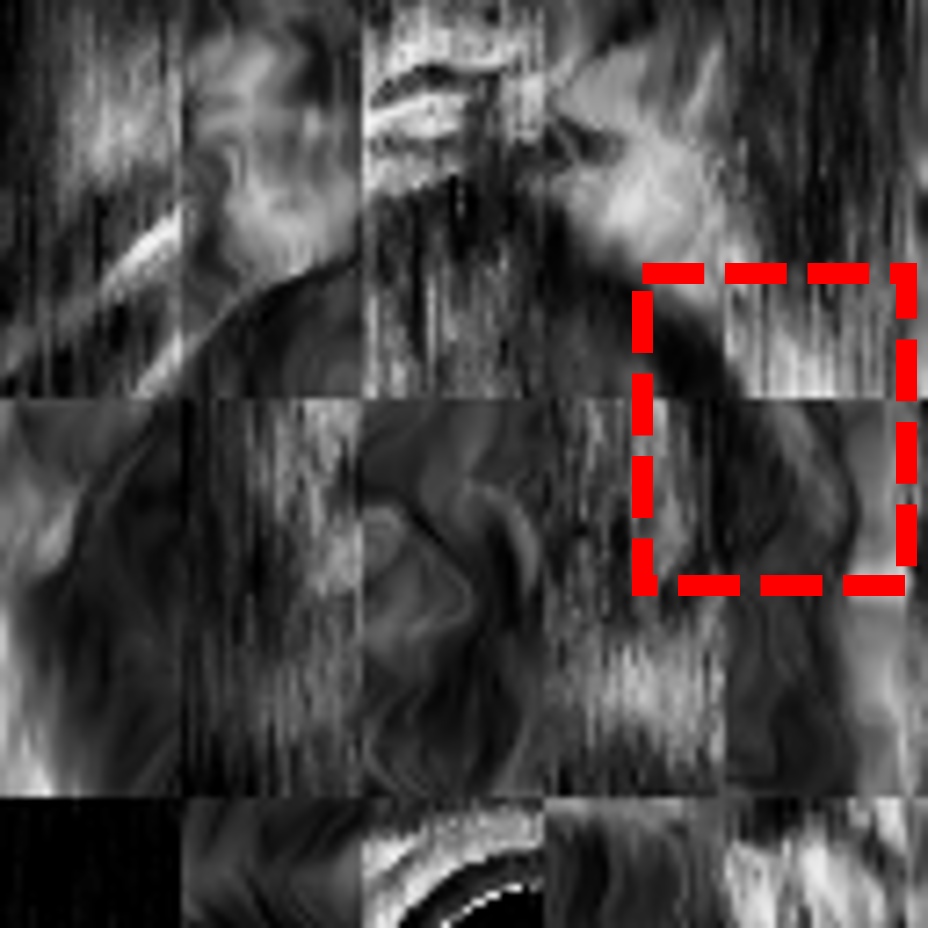}
    \end{subfigure}
    \begin{subfigure}{0.117\textwidth}
        \centering
        \includegraphics[width=\textwidth]{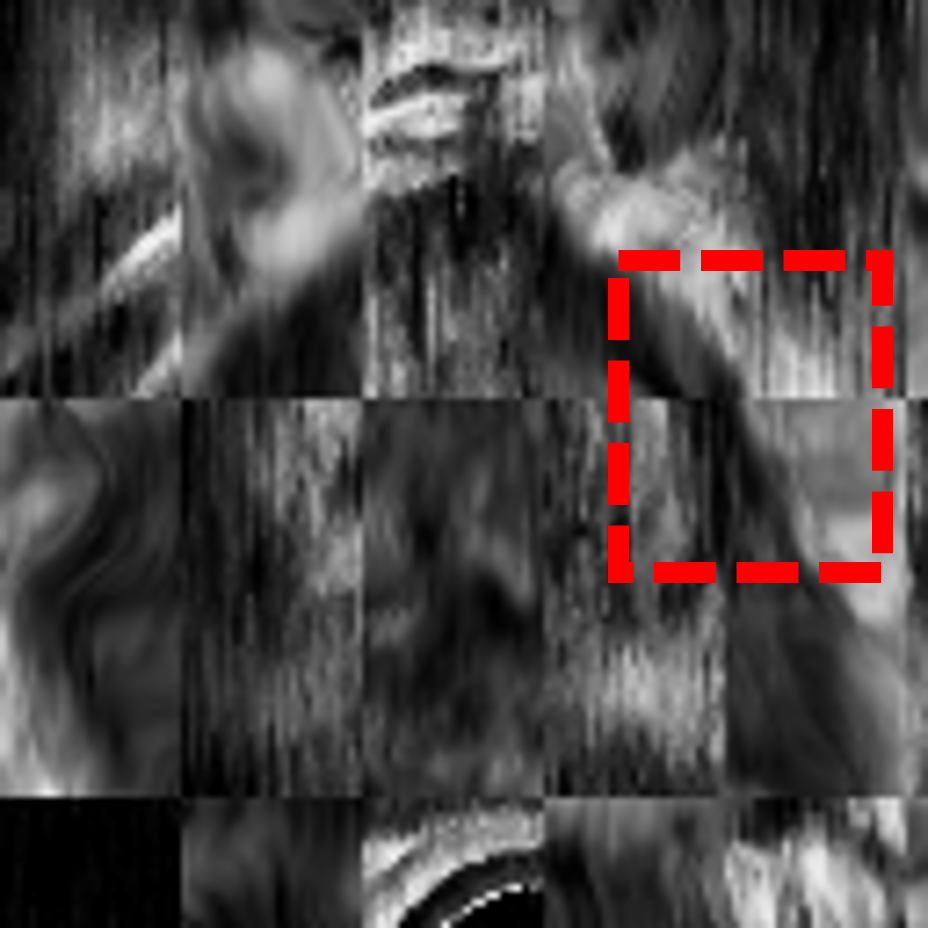}
    \end{subfigure}
    \begin{subfigure}{0.117\textwidth}
        \centering
        \includegraphics[width=\textwidth]{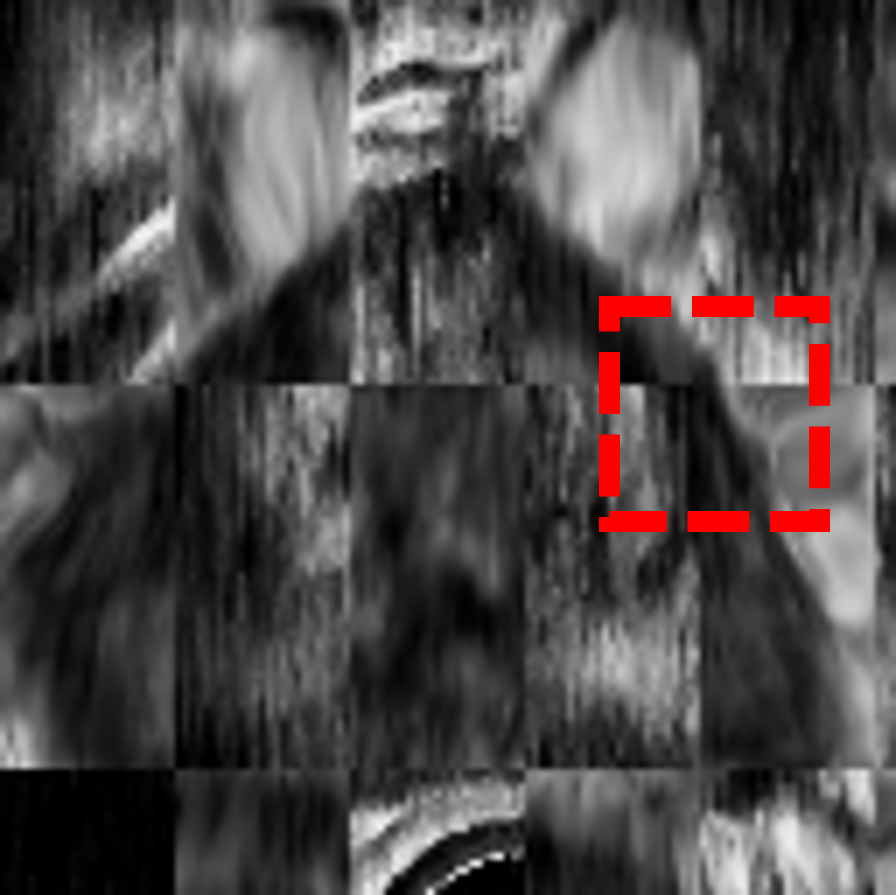}
    \end{subfigure}

    \caption{Registration results for two patients: The original US, original MR, and registration results using different modality translation methods, displayed in a chessboard pattern.}
    \label{fig:registration}
\end{figure*}

\subsubsection{Registration Performance}
To enable precise quantitative evaluation of registration performance, we conducted expert-guided manual segmentation of the prostate on several key frames of all test cases to serve as ground truth. The deformation fields generated by different methods were then directly applied to the binary segmentation masks of these key frames. The warped masks were subsequently compared with the manual ground truth to compute the following registration metrics:

\begin{itemize}
    \item \textbf{Dice Similarity Coefficient (DSC)}: 
    $\frac{2|X \cap Y|}{|X| + |Y|}$ measures volume overlap between registered and target masks (higher is better).
    
    \item \textbf{Intersection-over-Union (IoU)}: 
    $\frac{|X \cap Y|}{|X \cup Y|}$ provides stricter boundary alignment assessment (higher is better).
    
    \item \textbf{Average Surface Distance (ASD)}: 
    Mean Euclidean distance between corresponding segmentation surfaces (lower is better).
\end{itemize}
\noindent Additionally, to quantitatively assess the smoothness of the deformation fields, we employed the \textit{harmonic energy} (HE) metric \cite{smith2004advances}, defined as the squared Frobenius norm of the deformation field's Laplacian:

\begin{equation}
    \mathcal{E}_{\text{harmonic}} = \|\nabla^2 \phi\|_F^2 = \sum_{i=1}^3 \|\nabla^2 \phi_i\|^2
\end{equation}

\noindent where $\phi = (\phi_1, \phi_2, \phi_3)$ represents the 3D deformation field and $\nabla^2$ denotes the Laplace operator. This metric directly penalizes rapid spatial variations in the deformation by measuring its second-order derivatives, with lower values indicating smoother, more physically plausible transformations. 

\begin{table}[t]
    \centering
    \caption{Quantitative comparison of registration performance across different methods.}
    \label{tab:metrics}
    \begin{tabular}{lcccc}
        \toprule
        Method & DSC $\uparrow$ & IoU $\uparrow$ & ASD $\downarrow$ & HE $\downarrow$ \\
        \midrule
        DiffuseMorph & 0.88 & 0.79 & 12.66 & \num{9.07e4} \\
        FSDiffReg & 0.92 & 0.87 & 10.74 & \num{4.10e5} \\ \hline
        UNSB+FSDiffReg & 0.92 & 0.88 & 12.83 & \num{8.10e5} \\
        PMT+FSDiffReg & \textcolor{blue}{0.95} & \textcolor{blue}{0.91} & 9.18 & \num{4.54e5} \\
        ACMT+FSDiffReg & \textcolor{blue}{0.95} & 0.90 & \textcolor{blue}{6.82} & \textcolor{blue}{\num{2.60e5}} \\ \hline
        MR2US-Pro (ours) & \textbf{0.97} & \textbf{0.94} & \textbf{4.45} & $\mathbf{7.35 \times 10^{4}}$  \\
        \bottomrule
    \end{tabular}
\end{table}


    
    

The experimental results are shown in Table~\ref{tab:metrics}, which clearly demonstrate that our method (MR2US-Pro) achieves the best performance across all metrics. We provide a systematic analysis of the results as follows.

\begin{enumerate}
    \item \textbf{End-to-end Performance}: Our method demonstrates superior performance across all metrics compared to SOTA approaches (DiffusionMorph and FSdiffReg). In particular, existing methods struggle to reduce average surface distance (ASD), a key metric for evaluating prostate surface alignment in clinical use. In contrast, our method reduces ASD by a factor of three compared to DiffusionMorph, achieving a mean surface error of just 4 pixels, while others typically exceed 10 pixels. Such precision is essential in clinical practice, where even sub-millimeter discrepancies in anatomical alignment can have a profound impact on diagnostic accuracy and the precision of radiation therapy targeting.

    \item \textbf{Impact of Modality Translation}: When evaluating with FSdiffReg as the fixed registration backbone and varying only the translation modules, our ACMT approach consistently outperforms the alternatives, achieving the highest scores in Dice, ASD, and HE (blue colored in table \ref{tab:metrics}). Despite a marginal 0.01 IoU difference compared to PMT+FSdiffReg, our method stands out by reducing ASD by 25.7\% relative to the second-best performer (PMT+FSdiffReg), highlighting the exceptional surface alignment precision enabled by our modality translation component. These results validate that ACMT generates anatomically optimal intermediate representations, significantly improving registration accuracy when compared to UNSB and PMT alternatives.

    \item \textbf{Impact of Registration Architecture}: When fixing our ACMT for modality translation and comparing registration strategies, our complete framework (MR2US-Pro) outperforms ACMT+FSdiffReg across all evaluation metrics. Most notably, it reduces Harmonic Energy by an order of magnitude. These results provide conclusive evidence that our anatomy-aware registration strategy makes essential contributions beyond modality translation alone, with the significantly lower HE values particularly demonstrating its ability to generate more physically plausible deformation fields—an essential advantage for clinical applications requiring anatomically faithful registration results.
\end{enumerate}

\begin{figure*}[t]
    \centering

    \begin{subfigure}{0.15\textwidth}
        \centering
        \textbf{DiffuseMorph}
    \end{subfigure}
    \begin{subfigure}{0.15\textwidth}
        \centering
        \textbf{FSDiffReg}
    \end{subfigure}
    \begin{subfigure}{0.15\textwidth}
        \centering
        \textbf{UNSB+FSDiffReg}
    \end{subfigure}
    \begin{subfigure}{0.15\textwidth}
        \centering
        \textbf{PMT+FSDiffReg}
    \end{subfigure}
    \begin{subfigure}{0.15\textwidth}
        \centering
        \textbf{ACMT+FSDiffReg}
    \end{subfigure}
    \begin{subfigure}{0.15\textwidth}
        \centering
        \textbf{MR2US-Pro (Ours)}
    \end{subfigure}

    \vspace{0.5em} 

    \begin{subfigure}{0.15\textwidth}
        \includegraphics[width=\textwidth]{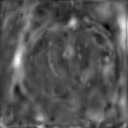}
    \end{subfigure}
    \begin{subfigure}{0.15\textwidth}
        \includegraphics[width=\textwidth]{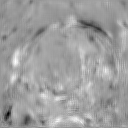}
    \end{subfigure}
    \begin{subfigure}{0.15\textwidth}
        \includegraphics[width=\textwidth]{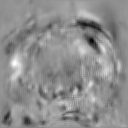}
    \end{subfigure}  
    \begin{subfigure}{0.15\textwidth}
        \includegraphics[width=\textwidth]{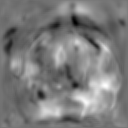}
    \end{subfigure}
    \begin{subfigure}{0.15\textwidth}
        \includegraphics[width=\textwidth]{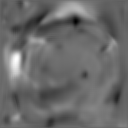}
    \end{subfigure}
    \begin{subfigure}{0.15\textwidth}
        \includegraphics[width=\textwidth]{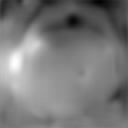}
    \end{subfigure}

    \vspace{0.1em} 

    \begin{subfigure}{0.15\textwidth}
        \includegraphics[width=\textwidth]{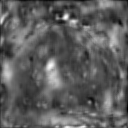}
    \end{subfigure}
    \begin{subfigure}{0.15\textwidth}
        \includegraphics[width=\textwidth]{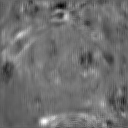}
    \end{subfigure}
    \begin{subfigure}{0.15\textwidth}
        \includegraphics[width=\textwidth]{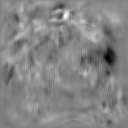}
    \end{subfigure}  
    \begin{subfigure}{0.15\textwidth}
        \includegraphics[width=\textwidth]{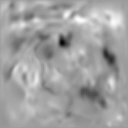}
    \end{subfigure}
    \begin{subfigure}{0.15\textwidth}
        \includegraphics[width=\textwidth]{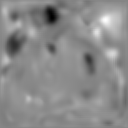}
    \end{subfigure}
    \begin{subfigure}{0.15\textwidth}
        \includegraphics[width=\textwidth]{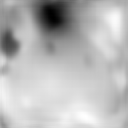}
    \end{subfigure}

    \caption{Deformation fields of different methods for two patients. Each row corresponds to one patient, and each column corresponds to one method.}
    \label{fig:deformation_field}
\end{figure*}

The visual results in Figure~\ref{fig:registration} further support the conclusions from our quantitative evaluation. In particular, in the red dashed boxes of Patient 2 (second row), only two methods achieve visibly accurate alignment: our full MR2US-Pro pipeline (which includes both our ACMT and the Anatomy-aware registration), and the combination of ACMT with FSDiffReg. All other methods exhibit evident structural discontinuities in this region, indicating registration failure. This highlights the importance of our modality translation strategy in enabling robust registration under difficult anatomical conditions.

In the case of Patient 1 (first row), the red dashed box shows that all modality translation methods, when combined with the same registration model (FSDiffReg), successfully improve alignment in a region where registration without translation fails. However, the yellow dashed box reveals another key distinction: while other methods force alignment between the fine MR boundary and the thicker US boundary, resulting in unnaturally thickened MR contours and visibly non-uniform deformation, only our MR2US-Pro method achieves a smoother, anatomically reasonable transformation. It avoids overfitting to high-information regions and instead places more emphasis on the unified low-information areas of the prostate interior, resulting in a natural transition in the boundary region without distorting anatomical features. This highlights the advantage of our anatomy-aware registration method, which effectively preserves anatomical coherence while aligning the image.

This observation is further validated by the deformation fields shown in Figure~\ref{fig:deformation_field}, where our method demonstrates significantly smoother transformations compared to others. Together, these visual cues underscore the superior robustness and anatomical fidelity of our proposed MR2US-Pro framework, especially in complex clinical scenarios where conventional approaches often struggle.

In summary, the proposed framework demonstrates consistent improvements from both the perspective of cross-modality translation and final registration performance. The synergy between anatomically coherent translation and anatomy-aware registration contributes to a more robust and clinically reliable pipeline for MR-US alignment. These results highlight the practical value of our design choices in addressing the challenges of cross-modality registration.

\section{Conclusion}
In this paper, we proposed a framework based on modality translation to address the challenges of cross-dimensional and cross-modal registration between Prostate MR and US images. To overcome the dimensional discrepancy, we first introduce a completely probe-location-independent TRUS 3D reconstruction method to convert 2D ultrasound sequences into dense 3D volumes. Subsequently, we leverage an ACMT network to transform both MR and US 3D volumes into a customized unified intermediate representation. This intermediate modality ensures highly anatomical consistency within the prostate while preserving important boundary features. Finally, we design an anatomy-aware registration method that focuses the alignment process on the anatomically consistent internal regions of the prostate, thereby enhancing registration accuracy and robustness.

Extensive quantitative and qualitative evaluations demonstrate that our framework outperforms existing state-of-the-art techniques in both modality translation and registration performance. Additionally, the core principle of our customized pseudo-modality translation makes the approach readily adaptable to a broad spectrum of cross-modal image alignment tasks, such as brain CT–MR and cardiac PET–MR registration.

\bibliographystyle{splncs04}
\bibliography{mybibliography}
\end{document}